\documentclass[a4paper,twocolumn,11pt,unpublished]{quantumarticle} %onecolumn
\pdfoutput=1
\usepackage[english]{babel}

\usepackage[letterpaper,top=2cm,bottom=2cm,left=3cm,right=3cm,marginparwidth=1.75cm]{geometry}

\usepackage{quantikz}
\usepackage{adjustbox}
\usepackage{pifont}

\usepackage{amsmath}
\usepackage{graphicx}
\usepackage{subcaption}
\usepackage{relsize}
\usepackage[colorlinks=true, allcolors=blue]{hyperref}
\usepackage[ruled,vlined]{algorithm2e}

\DeclareMathOperator{\Tr}{Tr}
\newcommand{\Bra}{\langle}
\newcommand{\Ket}{\rangle}

\title{Distributed Quantum Dynamics on near-term quantum processors}
\author{Vladyslav Bohun}
\author{Maxence Grandadam}
\affiliation{Haiqu, Inc., 95 Third Street, San Francisco, CA 94103, USA}
\author{Maciej Koch-Janusz}
\affiliation{Haiqu, Inc., 95 Third Street, San Francisco, CA 94103, USA}
\affiliation{Department of Physics, University of Z\"urich, 8057 Z\"urich, Switzerland}

\begin{document}
\maketitle

\begin{abstract}
Simulations of quantum dynamics are a key application of near term quantum computing, but are hindered by the twin challenges of noise and small device scale, which limit the executable circuit depths and the number of qubits the algorithm can be run on.
Towards overcoming these obstacles we develop and implement a distributed variant of the projected Variational Quantum Dynamics which we dub dp-VQD, which allows to simultaneously alleviate circuit depth and width limitations. We employ the wire cutting technique, which can be executed on the existing devices without quantum or classical communication.
We demonstrate the full variational training on noisy simulators, and execute and perform the reconstruction on real IBM quantum devices.
The algorithm allows to execute Hamiltonian evolution simulations for problem sizes exceeding devices' nominal qubit counts, and to combine multiple small devices in a distributed computation. We test our approach on the Heisenberg and Hubbard model dynamics.
\end{abstract}

\section{Introduction}
Quantum dynamics simulations are one of the main problem classes with the potential to demonstrate quantum advantage before the advent of full fault-tolerant quantum computing (FTQC). Their applications include quantum condensed matter and quantum chemistry \cite{Georgescu_2014_usecase_dynamics1,Kassal_2008_usecase_dynamics2,Chiesa_2019_usecase_dynamics3,shtanko2023uncoveringlocalintegrabilityquantum_usecase_dynamics4,arute2020observationseparateddynamicscharge_usecase_dynamics5,Nandkishore_2015_usecase_dynamics6,de_Jong_2022_usecase_dynamics7}, and classical problems mappable to Hamiltonian dynamics, such as combinatorial optimization \cite{farhi2000quantumcomputationadiabaticevolutionde_Jong_2022_usecase_combinatorics1,SantoroTosatti_2006_usecase_combinatorics2,herzog2024improvingquantumclassicaldecomposition_usecase_combinatorics3,makhanov2023quantumcomputingapplicationsflight_usecase_combinatorics4}, or computational fluid dynamics and partial differential equation solving \cite{leong2023variationalquantumsimulationpartial_usecase_pde1,Babbush_2023_usecase_pde2,sato2024hamiltoniansimulationtimeevolvingpartial_usecase_pde3,demirdjian2022variationalquantumsolutionsadvectiondiffusion_usecase_pde4,zamora2024efficientquantumlatticegas_usecase_pde5,hu2024quantumcircuitspartialdifferential_usecase_pde6,guseynov2024explicitgateconstructionblockencoding_usecase_pde7}. However, the depth of the quantum circuits in the simulations, typically employing Trotter methods \cite{PhysRevX.11.011020}, remains a major challenge, especially for longer times.

The difficulty is due to the existing Noisy Intermediate Scale Quantum (NISQ) devices suffering from severe limitations in both the width (qubit number) and depth (gate count) of the quantum circuits which can be accurately executed \cite{Nam_2019_noise_depth1,Childs_2018_noise_depth2,Pelofske_2022_qv1}.
The constraints are not independent: though state-of-art superconducting qubit quantum processors (QPUs) exceed $150$ qubits, the \emph{effective} number of qubits as measured be the logarithm of the Quantum Volume metric \cite{Pelofske_2022_qv1} remains an order of magnitude lower. Furthermore, even when error correcting codes (EC) are introduced allowing deeper simulations, the scaling limitations of the physical QPU platforms may restrict them to only a handful of logical qubits per chip. This calls for improved quantum dynamics algorithms, at least partially addressing hardware restrictions.

A number of approaches attempt to deal with the deep quantum dynamics circuits, including based on Counter-Adiabatic Driving \cite{demirplak2003adiabatic,PhysRevLett.104.063002,PhysRevLett.111.100502,Sels_2017}, Variational Fast Forwarding \cite{cirstoiu2020variational} and Quantum Krylov subspace methods \cite{Motta2019, Takahashi2025, Stair2020}. We extend the projected Variational Quantum Dynamics (p-VQD) algorithm \cite{Barison_2021_pvqd_original}, which is a quantum machine learning (QML) approach only executing quantum circuits of constant depth, at the expense of repeatedly performing variational optimization on the QPU, and the associated sampling overheads. The advantage of p-VQD is that it can be applied to time-dependent Hamiltoninans, and in principle allows distributed versions \cite{Gentinetta_2024_gatecutting_pvqd} -- we exploit both of these properties.

\begin{figure*}[!t]
    \centering
        \includegraphics[width=\textwidth]{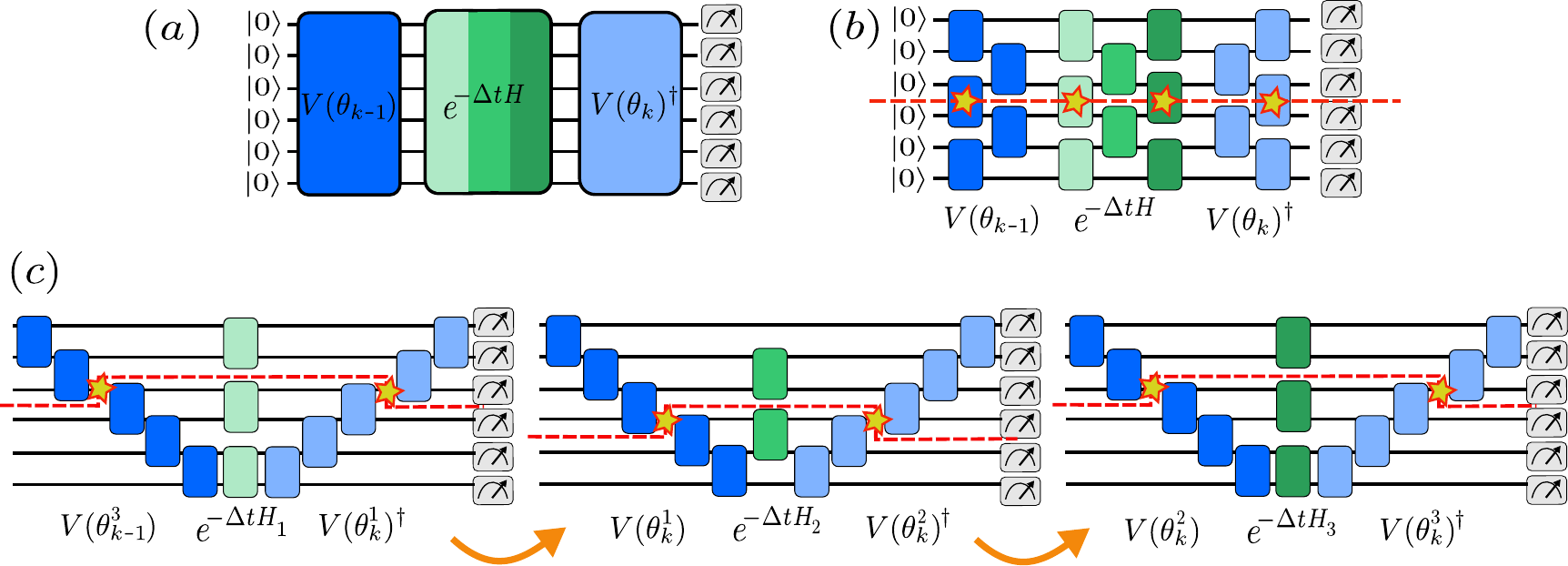}
        \caption{\textbf{(a)} The quantum circuit computing the loss Eq.\ref{eq:pvqd_stepk} in $k$-th iteration of the p-VQD procedure. Here $V(\theta_{k-1}^\gamma)$ with $\gamma=3$ is a PQC with parameters fixed in the previous iteration, and $V(\theta_{k}^\gamma)$ is the PQC currently trained using the Trotter step $\exp(-\Delta t H)$. The color shades on the Trotter operator represent its possible slicing, used in our algorithm (but not in p-VQD). \textbf{(b)} For a generic Trotter step the attempt to distribute p-VQD with circuit knitting will result in a large number of gate cuts (denoted by stars). Note that each cut circuit block may in fact contain more than one cut of elementary two-qubit gates. \textbf{(c)} Our dp-VQD algorithm uses sub-iterations of p-VQD, where each slice is constructed to be trivially cutable into two or more pieces, and correspondingly the PQC ansatz is chosen to be sparsly cutable in the same places (linear PQC is shown as an example), minimizing the total number of cuts. }
        \label{fig:computeuncompute}
\end{figure*}

Distributed quantum computing (DQC) allows to circumvent the scale limitations of QPU architectures \cite{Caleffi_2024}, and has also been proposed as an error mitigation technique \cite{Basu_2024_noise_reduction_cutting}. Importantly, DQC is also possible \emph{without} quantum communication between QPUs. On an algorithmic level methods known as circuit cutting/knitting \cite{Peng_2020_wire_cutting_original,Mitarai_2021_gate_cutting,Mitarai_2021_gate_cutting2,Lowe_2023_cutting_with_cc,Piveteau_2024_gate_cutting_with_cc} decompose a quantum circuit into a set of smaller ones which fit the QPU constraints, and whose combined results are used to reconstruct the outcomes of the original circuit, at the cost of increased sampling. These methods can be further distinguished by whether or not they require \emph{real-time classical} communication. Additionally, all gate cutting protocols require mid-circuit measurements (even if no communication is involved) also introducing additional noise, which is relatively large \cite{koh2024readout_midcirc_noise1,hines2024pauli_midcirc_noise2,hothem2024measuringerrorratesmidcircuit_midcirc_noise3}. While proof-of-principle experimental realization of classical communication aided DQC has recently been shown \cite{carrera2024combining}, here we focus on DQC without communication; our approach can easily be extended to the former case.

Here we combine both variational dynamics and distributed methods to end-to-end execute quantum dynamics simulations of simultaneously a larger depth, and utilising more qubits (physical or effective) than available on any single QPU. 
To this end we extend and modify p-VQD to efficiently merge it with the wire cutting protocol without classical communication (Sec.~\ref{sec:method}). The latter choice does not require mid-circuit measurements, and may even be executed off-line using a single device, allowing to execute the procedure end-to-end on the current NISQ processors. It furthermore suffers in practice from smaller sampling overheads and allows to simulate more complex Hamiltonians than possible with gate cutting, which we demonstrate by performing distributed Heisenberg and Hubbard model simulations on noisy simulators and actual IBM devices (see Sec.\ref{sec:experiments}).

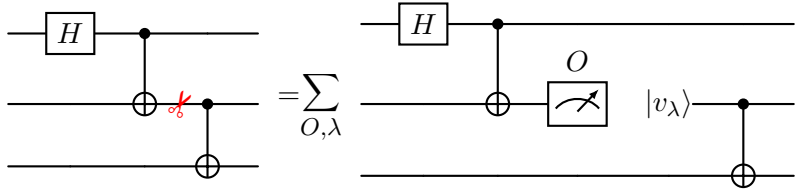
\begin{figure*}[!ht]
    \centering
    \begin{quantikz}
& \gate{H} & \ctrl{1} & &\\
&  & \targ{} \rstick{\color{red}\rotatebox[origin=c]{60}{\ding{34}}} & \ctrl{1} &\\
&  & & \targ{} &
\end{quantikz}
=$\mathlarger{\mathlarger{\mathlarger{\sum}}}_{O,\lambda}$
\begin{quantikz}
& \gate{H} & \ctrl{1} & & & &\\
&  & \targ{} &\meter{O} & \ket{v_\lambda}\wireoverride{n} & \ctrl{1} &\\
&  & & & & \targ{} &
\end{quantikz}
    \caption{A single cut (in red) of the original 3-qubit circuit splits it into disjoint 2-qubit subcircuits, which can be executed independently and whose results allow to reconstruct the outcome of the larger one.}
    \label{fig:cutting_example}
\end{figure*}

\section{Background}\label{sec:preliminaries}

\subsection{Projected Variational Quantum Dynamics}

The standard approach to simulating quantum dynamics is based on the Trotter approach: for a time-independent Hamiltonian $H = \sum_a H_a$, and an initial quantum state $|\psi_0\Ket$, the time evolved state at $t>0$ is approximated as:
\begin{equation}
    |\psi_t\Ket = e^{-i t H} |\psi_0\Ket \approx \prod_{j=1}^N e^{-i \Delta t H} |\psi_0\Ket,
\end{equation}
where $\Delta t = t / N$, and where to each operator $e^{-i \Delta t H}$ a Lie-Trotter-Suzuki decomposition is applied (\emph{e.g.}~to lowest order $\exp[ -i \Delta t H] \approx \prod_a \exp[ -i \Delta t H_a]$), approximating it with a product of evolutions containing only commuting terms. This allows to express the evolution as a quantum circuit. The approximation error depends on the time step $\Delta t$ and on the order of the decomposition; the method also applies to time-dependent Hamiltonians \cite{PhysRevX.11.011020}.

Projected Variational Quantum Dynamics (p-VQD)\cite{Barison_2021_pvqd_original} aims to address the
the fundamental problem of the linear growth of evolution circuits with simulation time, which in practice limits the accurate execution on current QPUs to a handful of Trotter steps. In this approach a parametrized quantum circuit (PQC) $V(\theta_k)$ with variational parameters $\theta_k$ approximates the quantum state $|\psi_k\Ket$ evolved to time $t_k = k\Delta t$:
\begin{equation}\label{eq:pqc_approx_definition}
    V(\theta_k) |0\Ket \approx  |\psi_k\Ket := |\psi_{k \Delta t}\Ket.
\end{equation}
The parameters $\theta_k$ are found by sequentially  maximizing the state fidelity between the state $|\psi_{k-1}\Ket$ evolved through the next Trotter step and the new approximation $V(\theta_k)|0\Ket$ of $|\psi_k\Ket$:
\begin{equation}
    \label{eq:pvqd_stepk}
    \theta_k = \arg \max_\theta |\Bra 0 | V(\theta)^\dagger e^{-i \Delta t H} V(\theta_{k-1}) | 0 \Ket|^2.
\end{equation}
Note that the circuit depth in Eq.\ref{eq:pvqd_stepk} is constant in each step.
The error depends both on the Trotter approximation and the fidelity of each variational training step (itself influenced by the PQC expressivity, and the device noise, among others), potentially accumulating. Nevertheless, for small enough time steps $\Delta t$ the procedure remains trainable, and barren plateau phenomena can be alleviated through the use of previous steps' parameters in initialisation \cite{Cerezo_2021_trainability2, puigivalls2024variational_warm_start}. Crucially, in practice the method allows the execution of longer  time dynamics than possible otherwise \cite{Barison_2021_pvqd_original}. It is also compatible with distributed approaches: in Ref.\cite{Gentinetta_2024_gatecutting_pvqd} the expressibility of a procedure combining p-VQD with gate-cutting was studied with constrained sampling overheads and in the absence of device noise.

The choice of the Ansatz is application dependent and may even be adaptive \cite{PhysRevResearch.6.023130}. We will use a fixed one (and thus of constant depth), whose form will be dictated by compatibility with the wire cutting procedure (see Sec.\ref{sec:method}).

\subsection{Wire cutting}
The wire-, or circuit-, cutting approach \cite{Peng_2020_wire_cutting_original,Lowe_2023_cutting_with_cc} is based on the following simple identity:
\begin{equation}\label{eq:wire_cutting_qpd}
    \rho = \sum_{\substack{O\in \mathcal{P}, \\ \lambda\in\sigma(O)}} \frac{\lambda}{2} \Tr(O\rho) \, |v_\lambda(O)\Ket\Bra v_\lambda(O)|,
\end{equation}
where $\rho$ is a 1-qubit density matrix, $\mathcal{P}=\{I, X, Y, Z\}$ are the Pauli matrices, $\sigma(O) \subseteq \{-1,1\}$ their eigenvalues, and $v_\lambda(O)$ the corresponding eigenvector. Eq.\ref{eq:wire_cutting_qpd} can be understood as decomposing the identity channel into a set of measurements in a particular basis, followed by a re-preparation in the corresponding eigenstate.

The identity Eq.\ref{eq:wire_cutting_qpd} can be applied to any qubit wire in a circuit, which is referred to as a \emph{cut}. The original circuit is decomposed into a weighted sum, whose terms are a product of two independent circuits (see Fig.\ref{fig:cutting_example}), each of which is a subcircuit with either a measurement or a state preparation operation inserted. Computations of expectation values of observables using the original circuit can instead be performed by executing a set of computations on the subcircuits, and combining their results according to Eq.\ref{eq:wire_cutting_qpd}. This does not require communication: all subcircuits can be executed in parallel (though classical communication versions of the procedure exist \cite{Lowe_2023_cutting_with_cc}). Details of the reconstruction procedure can be found in Refs.\cite{Tang_2021_cutqc,chen2023efficient_reconstruction_example}.

Circuit cutting comes at a price. First, there is the \textit{sampling overhead}: to estimate an observable to the same accuracy from a sum of observables via Eq.\ref{eq:wire_cutting_qpd} an overhead in measurement shots is needed which scales \emph{exponentially} in the number of cuts $k$. For the cutting procedure we use the overhead is $\mathcal{O}(\beta^{2k})$ with $\beta=4$
\cite{Mitarai_2021_overhead}.
Second, there is also a classical computational overhead: the sheer number of subcircuits, or, equivalently, the number terms in the reconstruction to be summed, scales exponentially in the number of cuts, making a large number of cuts infeasible. While reconstruction can be improved with tensor network methods, the worst-case estimates are still exponential \cite{tang2022scaleqc_scaleqc,mohseni2025buildquantumsupercomputerscaling}.

Due to these severe overheads it is essential to keep the number of cuts to a minimum, by either finding problems admitting a smart and sparse cut placement, or in \emph{e.g.~}QML applications by designing \emph{cuttable} Ans\"atze \cite{tomesh2023divide_vqe_example1,Marshall_2023_vqe_example2}. In what follows we show that certain important Trotterized Hamiltonian dynamics simulations display a sparsly-cuttable structure.

\section{The dp-VQD algorithm}\label{sec:method}

Here we combine the methods described in the previous section in a distributed projected variational quantum dynamics (dp-VQD) algorithm. Our strategy consists of two parts: first, we effectively \emph{create} a sparsly cuttable operator in the p-VQD by splitting a single Trotter step into multiple p-VQD \emph{sub-iterations}. Second, we simultaneously utilise a sparsly cuttable PQC ansatz, ensuring the cuttability of the whole procedure.

\begin{figure}[!t]
\centering
\includegraphics[width = 0.85\columnwidth]{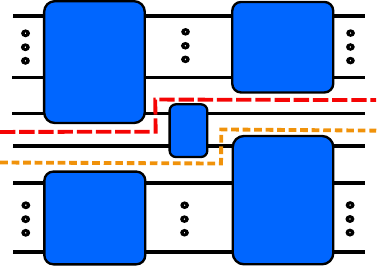}
    \caption{The structure of a PQC ansatz $V(\theta)$ which is cutable with a single cut on two adjacent wires. Each of the circuit subblocks, shown in blue, can have otherwise arbitrary structure and depth. The structure can equivalently be mirrored. This can be generalized to PQCs cuttable using $k>1$ cuts on selected subsets of qubits.}
    \label{fig:pqc_general}
\end{figure}

The p-VQD algorithm as introduced in Ref.\cite{Barison_2021_pvqd_original} absorbs a single complete Trotter step per iteration, as in Eq.\ref{eq:pvqd_stepk}. From the circuit cutting perspective, however, for a generic Hamiltonian the p-VQD iteration of full Trotter step is not easily cutable. Even for a simple Heisenberg Hamiltonian $\sum_i \mathbf{S}_i\mathbf{S}_{i+i}$, each Trotter-step circuit contains three layers of two-qubit gates implementing the Ising interactions for the $X,Y,Z$ Pauli operators (which already presupposes Ising interactions are in the native gate set). This requires three to six gate cuts, or one to three wire cuts, in addition to the cuts in the two copies of the ansatz $V(\theta)$ in Eq.\ref{eq:pvqd_stepk} (see Fig.\ref{fig:computeuncompute}b). With the overhead growing exponentially with the cut number, for a more complex Hamiltonian (potentially also necessitating a deeper Ansatz to support a lot of entanglement) this becomes unviable. 

To alleviate this problem we introduce p-VQD sub-iterations: the Trotter step circuit is subdivided into slices $T_\gamma$, each of which is easily cutable, preferably even trivially, and a p-VQD-like step is performed for these slices consecutively. Triviality here means that for each slice there exists a pair of qubits (not necessarily the same) or wires $j,j+1$ such that $T_\gamma \equiv T_\gamma^{1j}\otimes T_\gamma^{(j+1)n}$, \emph{i.e.}~the slice factorizes into a product of disconected circuits on wires $1\ldots j$ and $j+1 \ldots n$. For concreteness we focus on the case of cutability in two parts, but we emphasize that the discussion and the results trivially extend to multi-part cutability. Note that such a decomposition can always be found by simply taking a single layer of gates as multi-qubit (typically: 2-qubit) gates cannot be applied simultaneously to the same qubit. Depending on the properties of the Hamiltonian such decompositions with blocks $T_\gamma^{ab}$ of larger depth may exist, which can be exploited for more efficient sub-iterations (see Sec.\ref{sec:experiments} and Fig.\ref{fig:heisenberg_trotter} in the Appendix).

\begin{algorithm*}
    \SetKwInOut{Input}{Input}
    \SetKwInOut{Output}{Output}

    \Input{Hamiltonian $H$, number of trotter steps $N$, time step $\Delta t$ and the initial state $|\psi_0\rangle$}
    \Output{A sparsely cuttable circuit approximately preparing $|\psi_N\rangle$}

    $\triangleright$ $H$ is split into slices $T_\gamma$ ($\gamma=1\dots\Gamma$) s.t. each slice factorizes between one of two pairs of neighbouring qubits.\\

    $\triangleright$ A variational Ansatz $V(\theta)$ is prepared (see Fig.~\ref{fig:pqc_general}).

    $\triangleright$ Weights $\theta_0^\Gamma$ are randomly initialized.\\

    \For{$k = 1\dots N$}
    {

        \For{$\gamma=1\dots\Gamma$}
        {   
            \If{$\gamma = 1$}
            {
                \If{$k=1$}
                {
                    $|\psi_{\text{prev}}\rangle \gets |\psi_0\rangle$
                }
                \Else
                {
                    $|\psi_{\text{prev}}\rangle \gets V(\theta_{k-1}^\Gamma)|0\rangle$
                }
            }
            \Else
            {
                $|\psi_{\text{prev}}\rangle \gets V(\theta_{k}^{\gamma-1})|0\rangle$
            }
            
            $\theta_k^\gamma \gets \arg \max_\theta |\Bra 0 | V(\theta)^\dagger e^{-i \Delta t T_\gamma} | \psi_{\text{prev}} \Ket|^2$,\\
            where execution is done with wire cutting and the optimization is warm started from the previous weights.
        }
    }

    \Return{$V(\theta_N^\Gamma)$}
    
    \caption{dp-VQD algorithm.}\label{algo:dpvqd}
\end{algorithm*}

To take advantage of the trivial cutability of the slices created in each sub-iteration, we further build a matching Ansatz $V(\theta_k^\gamma)$. Here by matching we mean that the Ansatz requires only a few cuts on exactly the qubits defined by the factorized slices. As very simple example of this is shown in Fig.\ref{fig:computeuncompute}: a generic Trotter step circuit for \emph{e.g.}~the Heisenberg chain in Figs.\ref{fig:computeuncompute}a,b is split into three sub-iterations, where each $T_\gamma$ layer (depicted in shades of green) trivially factorizes between either second and third or third and fourth qubit (see Fig.\ref{fig:computeuncompute}c). A one-layer linear Ansatz can then be used in all of the sub-iterations, requiring only a single cut (thus two cuts in the full cost function computation Eq.\ref{eq:pvqd_stepk}). In the example in the figure the six qubit circuit can then be executed on smaller devices in three sub-iterations using two cuts each. Similar circuits have been considered for ground state search using variational quantum eigensolvers (VQE)\cite {tomesh2023divideconquercombinatorialoptimization,saleem2023approaches,khare2023parallelizingquantumclassicalworkloadsprofiling}. A more expressive Ansatz requiring the same amount of cuts is shown in Fig.\ref{fig:pqc_general}.

We emphasize that the choice of the linear Ansatz in Fig.\ref{fig:computeuncompute} was only to illustrate the cutability matching. In general more complex Ans\"atze are needed, whose structure and depth allow for sufficient expressivity and entanglement of different qubits \cite{Sim_2019_pqc1,dallairedemers2018lowdepth_pqc2,Kandala_2017_pqc3}. At the same time, with increased entangling capability, the number of required cuts necessarily grows. The exponential growth of resources with the number of cuts is a fundamental limitation of all circuit cutting/knitting methods. Here we do not overcome that constraint, but demonstrate that with the knowledge of the Hamiltonian, depending on its structure, the number of cuts can be efficiently minimized. A generic such situation may arise if the system can be decomposed into two or more \emph{weakly} entangled subparts, corresponding to \emph{e.g.}~a Hamiltonian of two weakly interacting parts of a molecule, each of them a strongly correlated system by itself. In Fig.\ref{fig:pqc_general} we show a more generic PQC structure requiring a single cut on two wires, and arbitrarily complex elsewhere. This can easily be generalized to families of circuits admitting $k$ cuts on designated qubits and otherwise arbitrary. Thus, both the structure of the sub-iterations and the Ansatz in dp-VQD are defined by the Hamiltonian (the Trotter step) to minimize the number of cuts.

\begin{figure*}[!t]
    \centering
    \begin{subfigure}[t]{0.49\textwidth}
    \centering
    \includegraphics[width=\textwidth]{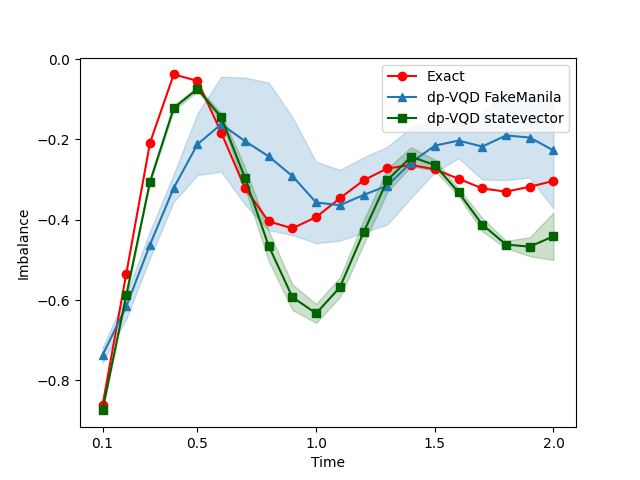}
    \caption{$w=1$}
    \label{fig:dp_vqd_manila_w1}
    \end{subfigure}
    \hfill
    \begin{subfigure}[t]{0.49\textwidth}
    \centering
    \includegraphics[width=\textwidth]{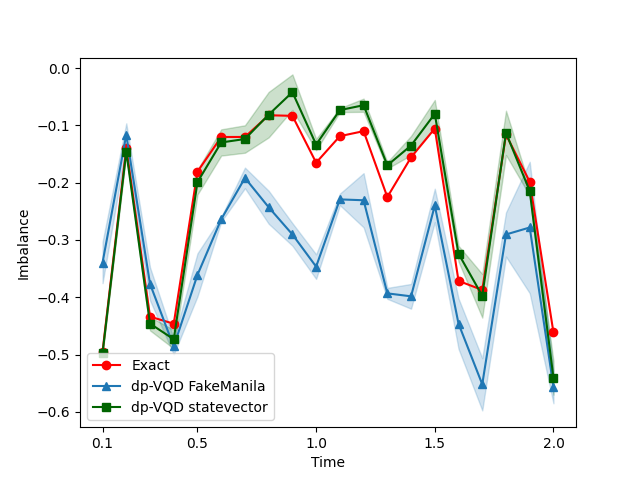}
    \caption{$w=10$}
    \label{fig:dp_vqd_manila_w10}
    \end{subfigure}
    \caption{Distributed simulation of time evolution of the 7-qubit disordered Heisenberg chain Eq.\ref{eq:heisenberg_hamiltonian} in the low $w=1$ \textbf{(a)} and high $w=10$ \textbf{(b)} disorder regimes, on the noisy IBM \texttt{FakeManila} 5-qubit backend. The evolution of the imbalance observable Eq.\ref{eq:imbalance_metric} is shown as a function of time. In red the exact noiseless simulation, in green, the dp-VQD algorithm on a noiseless simulator and in blue the dp-VQD algorithm on the noisy \texttt{FakeManila} backend. The dp-VQD algorithm executes only 5-qubit subcircuits, fitting the 5-qubit device constraint. The Trotter step is $\Delta t = 0.1$.}
    \label{fig:dp_vqd_manila_imbalance}
\end{figure*}

The circuit cutting allows to split the computation into circuit evaluation on fewer qubits, on either separate or the same devices performed asynchronously. The precise number of qubits necessary depends on the cut placement, itself dictated by the connectivity of the Trotter circuit, and whether qubit re-use \cite{pawar2023integrated_qubitreuse1,decross2022qubitreuse_qubitreuse2,fang2023dynamic_qubitreuse3} is performed. In the example of Fig.\ref{fig:computeuncompute} the six qubit circuit can be executed using five qubits (or four, with qubit re-use).

The complete variational training may be performed with sub-iterations using the loss Eq.\ref{eq:pvqd_stepk},  which we implement using the compute-uncompute method \cite{Havl_ek_2019_computeuncompute} (see Fig.\ref{fig:computeuncompute}a). Furthermore, one can modify the loss Eq.\ref{eq:pvqd_stepk}, which is a global function, to a local one, measuring the probability of return to the $| 0\rangle$ state on each qubit separately, as global loss functions inevitably lead to barren plateau (BP) phenomena, precluding training at scale \cite{Ragone_2024,larocca2024reviewbarrenplateausvariational}. Towards the same goal we use the previous iteration's weights as a warm start in PQC training \cite{puigivalls2024variational_warm_start}. Finally, we also take care to minimize the number of circuit evaluations by re-using partial subcircuit results in parameter updates (see App.\ref{sec:optimization}).

\section{Simulated and QPU results}\label{sec:experiments}

Here we apply the dp-VQD algorithm to execute in a distributed fashion two examples of quantum dynamics: the disordered Heisenberg chain, and the Hubbard chain with a weak link, on simulated and real IBM QPUs. In particular, we demonstrate the execution of quantum evolutions requiring more qubits than available on the device --  distributed computation is the only possibility to achieve this.

The disordered Heisenberg model on a 1D chain with open boundary conditions is given by:
\begin{equation}\label{eq:heisenberg_hamiltonian}
    H = J \sum_{i=1}^{n-1} \mathbf{S}_i\mathbf{S}_{i+1} + w \sum_{i=1}^{n} \left( h_{i,x} X_i + h_{i,z} Z_i \right),
\end{equation}
where $\mathbf{S_i} = (X_i,Y_i,Z_i)$ is the vector of Pauli operators on site $i$, $J$ and $w$ are the interaction and disorder strengths, respectively. The random disorder fields $h_{i,x}$ and $h_{i,z}$ are independent and identically distributed random variables from the uniform distribution on $[-1,1]$. The model is fundamental in the study of many-body localization (MBL) and displays a transition from a delocalized phase with ballistic entanglement spread at low disorder, to an MBL phase at high disorder \cite{PhysRevB.77.064426,RevModPhys.91.021001}. In our simulations we use $n=7$ qubits starting from an antiferromagnetically ordered state $|\dots 10101\Ket$. We use $J=1$, $\Delta t=0.1$ and vary disorder strengths $w$. The Trotter circuit for the evolution is given in Fig.~\ref{fig:heisenberg_trotter} in Appendix~\ref{app:heisenberg}: it can be split into two slices defining two dp-VQD sub-iterations. The corresponsing PQC ansatz is described in Appendix~\ref{app:pqc} and shown in Fig.~\ref{fig:our_pqc}.

\begin{figure*}[!t]
    \centering
    \begin{subfigure}[t]{0.49\textwidth}
    \centering
    \includegraphics[width=\textwidth]{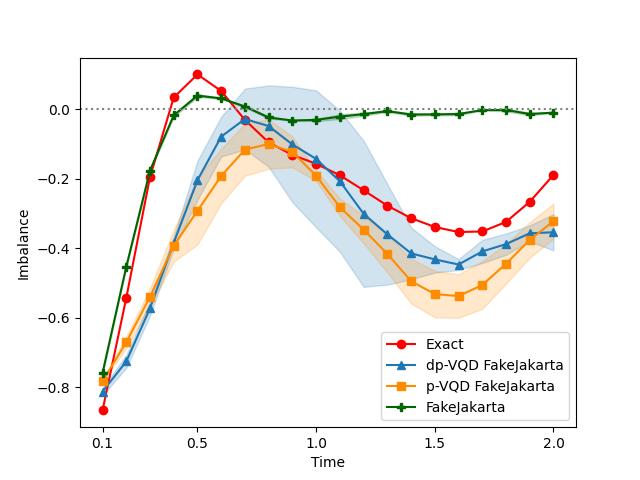}
    \caption{$w=1$}
    \label{fig:dp_vqd_jakarta_w1}
    \end{subfigure}
    \hfill
    \begin{subfigure}[t]{0.49\textwidth}
    \centering
    \includegraphics[width=\textwidth]{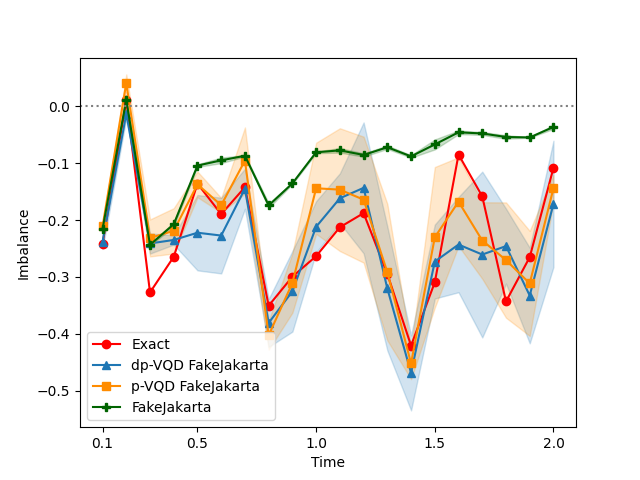}
    \caption{$w=10$}
    \label{fig:dp_vqd_jakarta_w10}
    \end{subfigure}
    \caption{Distributed simulation of time evolution of the 7-qubit disordered Heisenberg chain Eq.\ref{eq:heisenberg_hamiltonian} in the low $w=1$ \textbf{(a)} and high $w=10$ \textbf{(b)} disorder regimes, on the noisy IBM \texttt{FakeJakarta} 7-qubit backend. In contrast to Fig.\ref{fig:dp_vqd_manila_imbalance} the device allows for a direct execution of the 7-qubit Hamiltonian, whose results are shown in green: due to noise the imbalance observable decays very quickly to zero. This is in contrast to results from both our distributed (using only 5-qubit subcircuits) and the original p-VQD variants which are shown in orange and blue, respectively. These variational dynamics results reproduce the correct oscillatory dynamics of the exact (red) results for up to 20 steps. As in Fig.\ref{fig:dp_vqd_manila_imbalance}, the higher entanglement in the low disorder regime is more challenging for the algorithm and may require a more expressive ansatz to more accurately reproduce the observable dynamics.}
    \label{fig:dp_vqd_jakarta_imbalance}
\end{figure*}

%The Trotterization of this evolution is given in Appendix~\ref{app:heisenberg}. The input state for the evolution is considered an antiferromagnetic state, that is, the state $|\dots 10101\Ket$. In our experiments we set $J=1$, $\Delta t=0.1$, number of qubits $n=7$ and vary disorder parameter $w$. The detailed description of the PQC used in this and all following examples is given in Appendix~\ref{app:pqc}.

In Fig.~\ref{fig:dp_vqd_manila_imbalance} we show the results of executing the complete dp-VQD algorithm for the $n=7$ qubit model on the IBM \texttt{FakeManila} backend, a realistic noise model based on the \texttt{ibm\_manilla} 5-qubit QPU. We compare them to an an exact noiseless 7-qubit simulation of the full Hamiltonian, and a noiseless simulation of the distributed dp-VQD algorithm. The observable measured is the spin imbalance given by: 
\begin{equation}\label{eq:imbalance_metric}
    P = \sum_{i=1}^n (-1)^i Z_i.
\end{equation}

Both the noisy and noiseless dp-VQD results retain the qualitative features of the dynamics, \emph{i.e.}~the imbalance oscillations, for up to 20 Trotter steps, though the positions of the peaks are displaced due to the noise and the approximation errors in the variational procedure at the heart of p-VQD. Note that the results of a direct 7-qubit noisy execution are not shown, as the QPU does not have the requisite number of qubits -- a distributed approach is the only way to execute this problem on such a device. We observe that the dp-VQD results are closer to the exact ones in the large disorder regime. This is due to faster entanglement spreading at low disorder, which the single-cut weak link PQC Ansatz we used is unable to fully capture. This can be systematically improved by allowing a more expressive Ansatz requiring more cuts.

To obtain the results in Fig.~\ref{fig:dp_vqd_manila_imbalance} the variational procedure used 1024 shots per subcircuit to evaluate the loss (see the total shot budget in Fig.~\ref{fig:dp_vqd_manila_shots}), and the training was stopped when the loss plateaued. We note that the final losses (infidelity) on the noisy device were in the 0.3-0.5 range, corresponding to 0.02-0.15 on the ideal simulator (see Fig.\ref{fig:param_resilience}).

We repeated the same experiment on another simulated noisy QPU, the 7-qubit IBM \texttt{FakeJakarta}. The larger device allows to directly compare the distributed (which only executes 5-qubit subcircuits) with the regular variant of p-VQD, and with a direct execution of the original circuit in the presence of the same nontrivial noise model. As seen in Fig.~\ref{fig:dp_vqd_jakarta_imbalance} the direct execution results are quickly destroyed by the noise, with the imbalance observable decaying to zero. In contrast both variational procedures are able to reproduce the observable dynamics at longer times, for up to 20 Trotter steps. The distributed dp-VQD algorithm achieves similar results to the regular p-VQD execution, distributed computation coming at the cost of greater number of executions (of smaller circuits). %We note in passing that despite having a larger number of qubits, the \texttt{ibm\_jakarta} is in fact noisier than the smaller \texttt{ibm\_manilla} \cite{Pelofske_2022_qv1}.

Another area where simulations of quantum dynamics are particularly relevant and challenging is that of strongly correlated electronic systems. We test the execution of the dp-VQD evolution on the physical \texttt{ibm\_sherbrooke} QPU using the example of the paradigmatic Hubbard Hamiltonian. The model describes electrons hopping on a lattice, with a nearest-neighbour hopping $h$ and on-site interactions $U$:
\begin{equation}\label{eq:hubbard_ham}
    H = -\sum_{<i,j>, \sigma} h_{ij, \sigma} \left(c^{\dagger}_{i,\sigma} c_{j,\sigma} + \text{h.c}\right) + U \sum_{i} n_{i,\uparrow} n_{i\downarrow},
\end{equation}
where $c^{\dagger}_{i,\sigma}$ ($c_{i,\sigma}$) is the creation (destruction) operator for an electron with spin $\sigma$ on site $i$, and $n_{i,\sigma}$ counts the number of electrons with spin $\sigma$ on site $i$. While the ground-state properties of the Hubbard chain are extensively studied and much is known at least in certain limits \cite{Arovas2022}, its dynamical properties remain an open problem \cite{arute2020observationseparateddynamicscharge_usecase_dynamics5, Aoki2014}.

Here we specifically consider the example of a 1D chain of $6$ sites (thus $12$ qubits after mapping the fermions onto qubits) with $h_{ij, \sigma} = 1$, $U=5$ and a ``weak link'' \cite{Pati2004, Hettiarachchilage2013} where the hopping term for one spin species is set to zero ($h_{34, \downarrow} = 0$). It can be thought of as a simple model of a heterostructure with a spin-selective barrier between two correlated layers, such as a magnetic tunnel junction \cite{Meservey1994} or a magnetic Josephson junction in the normal state. This presents a natural use case for a sparsely cutable ansatz, mimicking the weak link structure of the Hamiltonian.

We study the dynamics of the staggered magnetization observable giving insight into antiferromagnetic correlations:
\begin{equation}\label{eq:stag_mag}
    m = \frac{1}{6} \left(m_1 - m_2 + m_3 - m_4 + m_5 - m_6\right),
\end{equation}
where the spin density at each fermionic site $1\leq i \leq 6$ is given by $m_i = (Z_{2i-1} - Z_{2i})/2$
The system is initialized in an antiferromagnetic state ($|100110011001\Ket$ in the qubit space) with $N/2$ spin-up and $N/2$ spin-down electrons on both sides of the weak link, and evolved for up $50$ Trotter steps with $\Delta t=0.3$ using the dp-VQD algorithm.
While the circuit representing each Trotter step is now much deeper than in the previous examples (45 gate layers per step), it still can be executed using two dp-VQD subiterations using a single cut (see App.~\ref{app:hubbard} for details).

\begin{figure}[!t]
    \centering
    \includegraphics[width = \columnwidth]{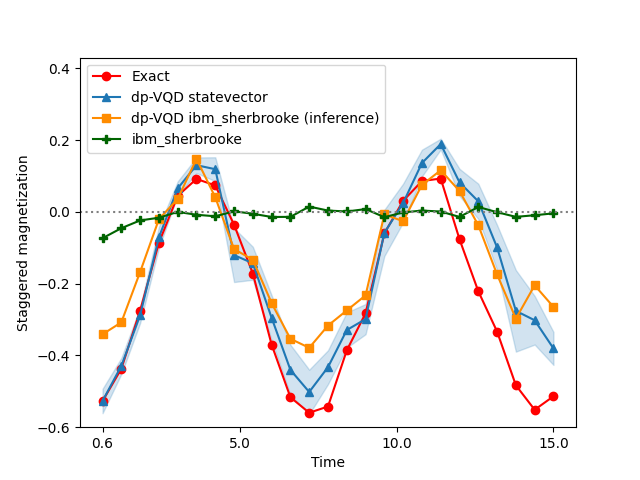}
    \caption{Distributed simulation of time evolution of a Hubbard model Eq.\ref{eq:hubbard_ham} with a spin-selective "weak-link" (see main text). The staggered magnetization observable Eq.\ref{eq:stag_mag} is simulated for up to 50 Trotter steps with $\Delta t=0.3$. In red the exact state vector simulator results for the complete 12-qubit Hamiltonian, in blue the dp-VQD algorithm using at most 8-qubit subcircuits executed on the exact simulator, in orange the dp-VQD algorithm trained on the exact simulator, but executed on the physical \texttt{ibm\_sherbrooke} QPU in a distributed manner using at most 7 qubits, and in green the direct execution of the full evolution circuit on the QPU.}
    \label{fig:dp_vqd_hubbard}
\end{figure}

In Fig.~\ref{fig:dp_vqd_hubbard} we show the results of the dp-VQD execution on the \texttt{ibm\_sherbrooke} quantum computer. Here, for cost reasons, the variational training is performed on an exact simulator, but the final subcircuits are executed in a distributed manner on the physical QPU, using no more than 7 qubits (for the 12 qubit Hamiltonian). As can be seen the results (orange) closely follow the exact staggered magnetization dynamics, whereas a direct execution of the standard Trotterized dynamics quantum circuits, shown in green in Fig.~\ref{fig:dp_vqd_hubbard}, is rapidly destroyed by the noise. In these proof-of-concept experiments we used 1024 shots per subcircuit and deployed no error mitigation (EM). The results demonstrate that distributed variational dynamics algorithms can extend the range of quantum dynamics simulations and are readily implementable already on the existing quantum hardware.

\section{Conclusions and outlook}\label{sec:conclusions}
In this work, motivated by the challenges of simulating quantum dynamics on NISQ devices, we introduced and implemented the distributed dp-VQD algorithm extending the p-VQD procedure \cite{Barison_2021_pvqd_original}, allowing to execute both deeper and wider circuits on the current quantum devices. Our technique combined the projected variational quantum dynamics approaches with circuit cutting and problem-aware PQC Ansatz construction. 

By executing in a distributed fashion Heisenberg and Hubbard model dynamics on the simulated and real noisy quantum devices, we demonstrated that our approach is directly applicable on the current QPUs and allows to extend the range of computation of quantum observables beyond that possible by directly running the Trotterized dynamics. We showed that quantum evolution circuits exceeding the nominal qubit count of devices can thus be executed: up to $m$-fold larger circuits when cutting the sub-iteration into $m$ pieces (in our experiments we used $m=2$).
Our technique will be relevant not only for the small devices we tested it on, but can be deployed to large ones limited by their effective Quantum Volume and not necessarily by the number of physical qubits (as is the case in \emph{e.g.}~the current state-of-art superconducting QPUs). 

In the implementation which we described we took care to optimize the resources used through a choice of parameter optimizers, Ans\"atze and cutting algorithm, allowing to minimize the number of necessary circuit evaluations, which due to the current high costs are of a significant practical concern. We also used the Rivet transpiler \cite{RivetGithub}, allowing to efficiently cash and re-use previously transpiled subcircuits during basis changes, saving classical computational resources.

Our approach can naturally be extended in multiple ways: first, variants of wire cutting with classical communication can be employed \cite{Lowe_2023_cutting_with_cc}. As mentioned in the main text, more complex sparsely cuttable Ans\"atze can be constructed, admitting precisely $k$ cuts, either at the same qubits, allowing for more entanglement, or in additional places, allowing to separate circuits into yet smaller parts. Both approaches can be combined and an optimal combination selected within a prescribed budget of cuts, the overhead of which is the main limiting factor of any cutting method. As cutting methods have been proposed to act effectively as an error mitigation (EM) \cite{Basu_2024_noise_reduction_cutting,khare2023parallelizingquantumclassicalworkloadsprofiling}, it will also be interesting to investigate the performance of dp-VQD in combination with standard EM tools.

We expect distributed quantum dynamics simulation methods to be important in taking advantage of the capabilities of the early quantum devices.

\bibliographystyle{quantum}
\bibliography{sts_distr}

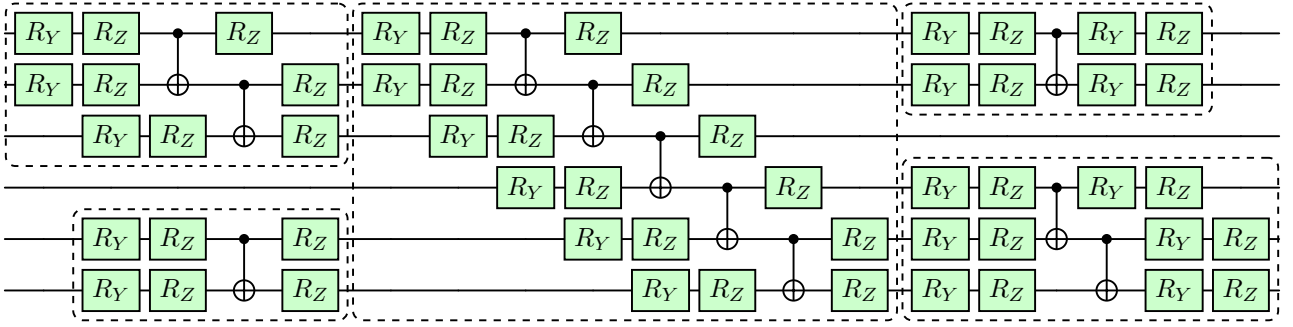
\begin{figure*}[!t]
    \centering
    \begin{adjustbox}{width=\textwidth}
    \begin{quantikz}[background color=green!20,column sep={0.15cm},row sep={0.15cm}]
&\gate{R_Y}\gategroup[3,steps=5,style={dashed,rounded
corners,inner
xsep=0pt,inner ysep=0pt},background]{}&\gate{R_Z}&\ctrl{1}&\gate{R_Z}&&[0.2cm]\gate{R_Y}\gategroup[6,steps=8,style={dashed,rounded
corners,inner
xsep=0pt,inner ysep=0pt},background]{}&\gate{R_Z}&\ctrl{1}&\gate{R_Z}&&&&&[0.2cm]\gate{R_Y}\gategroup[2,steps=5,style={dashed,rounded
corners,inner
xsep=0pt,inner ysep=0pt},background]{}&\gate{R_Z}&\ctrl{1}&\gate{R_Y}&\gate{R_Z}&&\\
&\gate{R_Y}&\gate{R_Z}&\targ{}&\ctrl{1}&\gate{R_Z}&\gate{R_Y}&\gate{R_Z}&\targ{}&\ctrl{1}&\gate{R_Z}&&&&\gate{R_Y}&\gate{R_Z}&\targ{}&\gate{R_Y}&\gate{R_Z}&&\\
&&\gate{R_Y}&\gate{R_Z}&\targ{}&\gate{R_Z}&&\gate{R_Y}&\gate{R_Z}&\targ{}&\ctrl{1}&\gate{R_Z}&&&&&&&&&\\
&&&&&&&&\gate{R_Y}&\gate{R_Z}&\targ{}&\ctrl{1}&\gate{R_Z}&&\gate{R_Y}\gategroup[3,steps=6,style={dashed,rounded
corners,inner
xsep=0pt,inner ysep=0pt},background]{}&\gate{R_Z}&\ctrl{1}&\gate{R_Y}&\gate{R_Z}&&\\
&&\gate{R_Y}\gategroup[2,steps=4,style={dashed,rounded
corners,inner
xsep=0pt,inner ysep=0pt},background]{}&\gate{R_Z}&\ctrl{1}&\gate{R_Z}&&&&\gate{R_Y}&\gate{R_Z}&\targ{}&\ctrl{1}&\gate{R_Z}&\gate{R_Y}&\gate{R_Z}&\targ{}&\ctrl{1}&\gate{R_Y}&\gate{R_Z}&\\
&&\gate{R_Y}&\gate{R_Z}&\targ{}&\gate{R_Z}&&&&&\gate{R_Y}&\gate{R_Z}&\targ{}&\gate{R_Z}&\gate{R_Y}&\gate{R_Z}&&\targ{}&\gate{R_Y}&\gate{R_Z}&
    \end{quantikz}
    \end{adjustbox}
\caption{A 6-qubit parametrized quantum circuit (PQC) whose layout corresponds to the general structure in Fig.~\ref{fig:pqc_general}.}
\label{fig:our_pqc}
\end{figure*}

\newpage
\appendix

\section{PQC Ansatz for dp-VQD}\label{app:pqc}

An example of a sparsely cutable PQC of the type we used in our experiments is shown in Fig.~\ref{fig:our_pqc} (for 6 qubits). All $R_Y$ and $R_Z$ gates are parametrized, with independent parameters. The structure can be expanded to any amount of qubits $n>3$, and the number of layers in corner blocks can be varied.

To implement the wire cutting in practice we use the method of Ref.\cite{Tang_2021_cutqc}, which, by grouping certain subcircuits, requires executing only four subcircuit pairs, as opposed to eight pairs in the naive implementation of Eq.\ref{eq:wire_cutting_qpd}.
While the sampling overhead is unchanged, it reduces the number of executed circuits and the classical overhead, which are important practical considerations, as QPU access is expensive in the NISQ era. As we restrict the number of allowed cuts to a constant factor (in the experiments: two), the full procedure has a constant overhead. This is in contrast to Ref.\cite{Gentinetta_2024_gatecutting_pvqd}, where the overhead depends on the amount of gates cut in a Trotter step (which grows with more complex Hamiltonians), and the values of their parameters.

\begin{figure}[!t]
    \centering
    \includegraphics[width=0.5\textwidth]{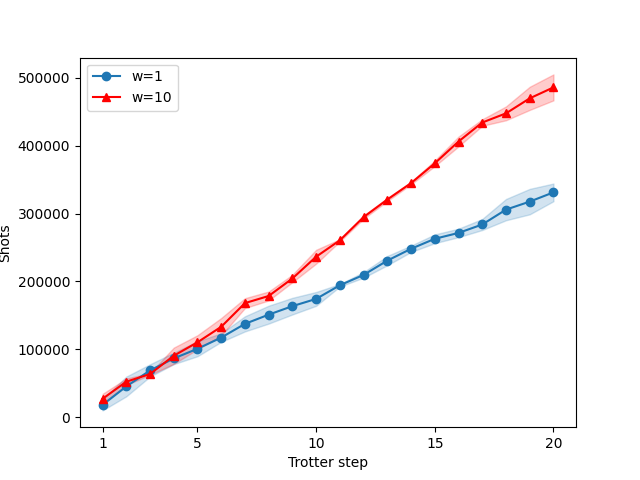}
    \caption{Shots spent on dp-VQD execution on the \texttt{FakeManila} device for different values of the disorder parameters $w$.}
    \label{fig:dp_vqd_manila_shots}
\end{figure}

\section{Optimization details}\label{sec:optimization}

We use the NFT optimizer \cite{Nakanishi_2020_nft} to minimize the cost function (infidelity). NFT is a gradient-free method exploiting periodicity properties of loss to instantly find an optimal value for a single parameter. In practice it shows a very good convergence rate, also in the presence of noise \cite{Nakanishi_2020_nft,Palackal_2023_nft_bench}. To use NFT each optimization parameter must correspond to a single rotation gate, the generator of which must be involutive, and the loss function must be a weighted sum of Hermitian observables with possibly different input states. These conditions are not very restrictive, and in particular are satisfied for our loss and Ansatz.

Using the NFT offers a computational advantage: for a single parameter update in NFT the loss is evaluated in several points with only this parameter changed. Consequently, since in the wire cutting procedure each subcircuit is computed independently, only one of them will be affected by the NFT parameter update. Thus, all evaluation results of the other subcircuits can be reused when the full loss value is reconstructed. This effectively reduces the amount of QPU evaluations during the optimization by a factor of two. This would also be true when using other parameter update methods, \emph{e.g.}~the parameter shift rule.

\begin{figure}[!t]
    \centering
    \includegraphics[width=0.5\textwidth]{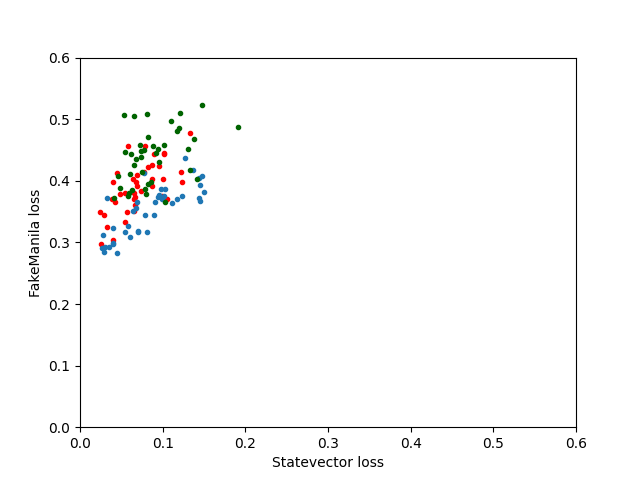}
    \caption{The final training loss (infidelity) of the dp-VQD subiterations for the disordered Heisenberg model Eq.\ref{eq:heisenberg_hamiltonian} evaluated on the simulated noisy device and on the ideal simulator, for three different random field configurations. Note the strong correlation between the loss values with and without noise.}
    \label{fig:param_resilience}
\end{figure}

\section{Heisenberg evolution implementation}\label{app:heisenberg}

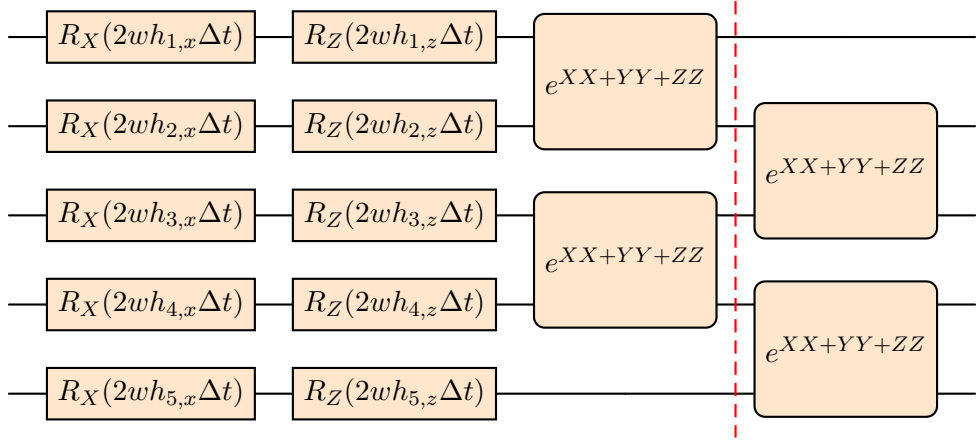
\begin{figure*}[t]
    \centering
    \begin{quantikz}[background color=orange!20]
&\gate{R_X(2wh_{1,x}\Delta t)}&\gate{R_Z(2wh_{1,z}\Delta t)}&\gate[2,style={rounded corners}]{e^{XX+YY+ZZ}}\slice{}&&\\
&\gate{R_X(2wh_{2,x}\Delta t)}&\gate{R_Z(2wh_{2,z}\Delta t)}&&\gate[2,style={rounded corners}]{e^{XX+YY+ZZ}}&\\
&\gate{R_X(2wh_{3,x}\Delta t)}&\gate{R_Z(2wh_{3,z}\Delta t)}&\gate[2,style={rounded corners}]{e^{XX+YY+ZZ}}&&\\
&\gate{R_X(2wh_{4,x}\Delta t)}&\gate{R_Z(2wh_{4,z}\Delta t)}&&\gate[2,style={rounded corners}]{e^{XX+YY+ZZ}}&\\
&\gate{R_X(2wh_{5,x}\Delta t)}&\gate{R_Z(2wh_{5,z}\Delta t)}&&&
    \end{quantikz}
\caption{A single Trotter step of Heisenberg evolution on $5$ qubits. The red dashed line splits it into two pieces comprising the two sub-iterations of dp-VQD with wire cutting.}
\label{fig:heisenberg_trotter}
\end{figure*}

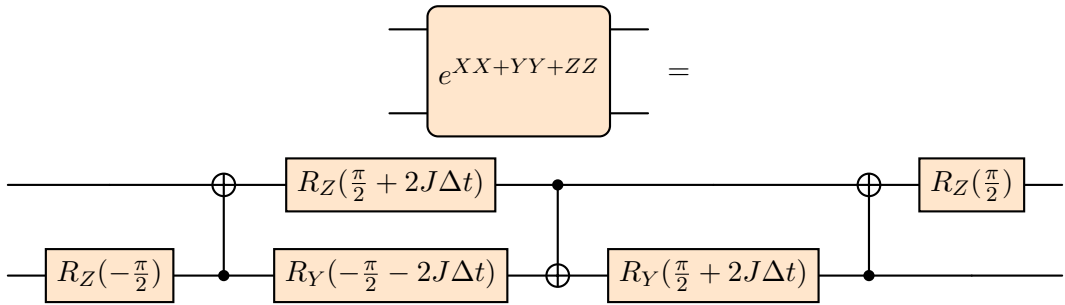
\begin{figure*}[t]
    \centering
    \begin{quantikz}[background color=orange!20]
&\gate[2,style={rounded corners}]{e^{XX+YY+ZZ}}&\\
&&
    \end{quantikz}
$=$
    \begin{quantikz}[background color=orange!20]
&&\targ{}&\gate{R_Z(\frac{\pi}{2} + 2J\Delta t)}&\ctrl{1}&&\targ{}&\gate{R_Z(\frac{\pi}{2})}&\\
&\gate{R_Z(-\frac{\pi}{2})}&\ctrl{-1}&\gate{R_Y(-\frac{\pi}{2}-2J\Delta t)}&\targ{}&\gate{R_Y(\frac{\pi}{2} + 2J\Delta t)}&\ctrl{-1}&&
    \end{quantikz}
\caption{Optimal $e^{XX+YY+ZZ}$ gate decomposition.}
\label{fig:xxyyzz}
\end{figure*}

A single Trotter step for the Hamiltonian evolution given by Eq.\ref{eq:heisenberg_hamiltonian} is shown in Fig.~\ref{fig:heisenberg_trotter} for $n=5$ qubits. The vertical line in the image separates the two slices which define the dp-VQD sub-iterations with wire cutting as in examples in Sec~\ref{sec:method}. A two-qubit block which implements $e^{XX+YY+ZZ}$ gate is shown in Fig.~\ref{fig:xxyyzz} and is its optimal decomposition (see \cite{Vatan_2004_xxyyzz}) in the number of \texttt{CNOT} gates. We note that it is also possible to write this operator as a sequence of $R_{XX}$, $R_{YY}$ and $R_{ZZ}$ rotations, however, their decomposition would require six \texttt{CNOT}s. The resulting Trotter step has depth 13, counted in layers of elementary gates.

\section{Hubbard model evolution implementation}\label{app:hubbard}

The Hubbard Hamiltonian Eq.\ref{eq:hubbard_ham} is a paradigmatic model in the study of electrons on a lattice subject to on-site interactions. It hosts a large variety of ordered phases due to the competition between the kinetic energy controlled by the hopping $h$, and the two-body interaction $U$ that tends to localize electrons. In 1D its ground state is a Luttinger liquid with separation of the spin and charge degrees of freedom. There are, however, still open questions regarding the fate of the Hubbard model when driven out of equilibrium. In our case we focused additionally on simualting the evolution in weak spin-selective tunneling scenario.

To this end we map the fermionic Hamiltonian \eqref{eq:hubbard_ham} to a qubit Hamiltonian using the Jordan-Wigner transformation. We study a chain with $N=6$ sites and we fix $h_{ij, \sigma} = 1$, for all $i,j $ except for $h_{34, \downarrow}=0$ and $U=5$. This allows us to generate a 12-qubit circuit with the following Pauli generators $G$: \textbf{(i)} $YZY$ and $XZX$ on all $i,i+i,i+2$ qubits except $i=5$ (counting from 1) with coefficient $-0.5$, \textbf{(ii)} $ZZ$ on $2i-1,2i$ qubits for $1\leq i \leq 6$ with coefficient $1.25$ and \textbf{(iii)} $Z$ for all qubits with coefficient $-1.25$.
For each mentioned Pauli generator $G$, we used the standard decomposition of $e^{-i \Delta t G}$ into gates. The depth of this Trotter step is 45, yet all gates can be easily arranged so that the dp-VQD procedure requires only two substeps.

\end{document}